\documentclass[twocolumn,final,sort&compress]{svjour3}                     
\smartqed  

\usepackage{graphicx}
\usepackage{amsmath}
\usepackage{amssymb}
\usepackage{dsfont}

\usepackage[numbers]{natbib}

      \newcommand{\conjg}[1]{\ensuremath{\hspace{1pt}\overline{\hspace{-1pt}#1\hspace{-1pt}}}\hspace{1pt}}

\def\Slash#1{\setbox0=\hbox{$#1$} 
\dimen0=\wd0 
\setbox1=\hbox{/} \dimen1=\wd1 
\ifdim\dimen0>\dimen1 
\rlap{\hbox to \dimen0{\hfil/\hfil}} 
#1 
\else 
\rlap{\hbox to \dimen1{\hfil$#1$\hfil}} 
/ 
\fi}

\begin{document}

\title{Delta properties in the rainbow-ladder truncation of Dyson-Schwinger equations
\thanks{"Relativistic Description of Two- and Three-Body Systems
in Nuclear Physics", ECT*, October 19-13 2009"}
}

\author{D.~Nicmorus         \and
        G.~Eichmann \and
        A.~Krassnigg \and
        R.~Alkofer
}

\institute{   D.~Nicmorus  \at
              Frankfurt Institute for Advanced Studies (FIAS),\\
              Johann Wolfgang Goethe-Universit\"{a}t, \\
              D-60438 Frankfurt am Main, Germany\\
              \email{nicmorus@th.physik.uni-frankfurt.de}            \\
              \emph{Present address: \\
              Thomas Jefferson National Accelerator Facility, \\
              Newport News, VA 23606, USA }
              \and
              G.~Eichmann \at
              Institut f\"{u}r Kernphysik, Technische Universit\"at Darmstadt, \\
              D-64289 Darmstadt, Germany
              \and
              A.~Krassnigg \and R.~Alkofer \at
              Institut f\"ur Physik, Karl-Franzens-Universit\"at Graz, \\
              A-8010 Graz, Austria
              }

\maketitle

\begin{abstract}

        We present a calculation of the three-quark core contribution to nucleon and $\Delta$-baryon masses
        and $\Delta$ electromagnetic form factors in a Poincar\'{e}-covariant Faddeev approach.
        A consistent setup for the dressed-quark propagator, the quark-quark, quark-'diquark'
        and quark-photon interactions is employed, where all ingredients are solutions of
        their respective Dyson-Schwinger or Bethe-Salpeter equations in a rainbow-ladder truncation.
        The resulting $\Delta$ electromagnetic form factors concur with present experimental and lattice data.


\end{abstract}


\section{Introduction}
\label{intro}

        The exploration of the rich structure of the nucleon represents one of the main tasks of
        contemporary particle physics. Present experimental facilities report accurate
        measurements of the nucleon's electromagnetic form factors. 
        The lowest-lying excited state of the nucleon, the $\Delta(1232)$ baryon, is produced 
        at energies above the pion-production threshold and
        plays an important role in nuclear strong interactions.
        A comprehensive study of $\Delta$-baryon properties in connection to those of the nucleon
        is expected to answer important issues of contemporary research such as the chiral cloud content
        of baryons and its impact on baryon properties. In this view, a comparative analysis of the $N\Delta\gamma$
        and the $\Delta \Delta \gamma$ transitions will elucidate the nature of the $\Delta$-baryon as a
        pure quark state, rather than a molecular state, and reveal connections between experimental observations and the
        fundamental phenomena that govern the physics of hadronic constituents.

        A quark-core analysis of nucleon and $\Delta$ masses and electromagnetic form factors
        has recently been carried out 
        in the Dyson-Schwinger approach~\cite{Nicmorus:2010sd,Nicmorus:2008vb,Nicmorus:2008eh,Eichmann:2008ef,Eichmann:2008kk}.
        Dynamical chiral symmetry breaking and confinement, two genuinely non-pertur\-bative phenomena
        tightly connected with the formation of bound states, can be consistently addressed
        only within a non-perturbative approach to QCD. 
        Such a framework is provided by the Dyson-Schwinger equations (DSEs) which are
        an infinite set of coupled integral equations for QCD's Green functions;
        see e.g.~\cite{Roberts:1994dr,Alkofer:2000wg,Fischer:2006ub} for reviews.
        In this context, 
        hadron properties are studied via covariant bound-state
        equations~\cite{Maris:2003vk,Roberts:2007jh,Eichmann:2009zx}: mesons ($q\bar{q}$ bound states)
        can be described by solutions of their Bethe-Salpeter equations (BSEs); baryons ($qqq$ bound states) 
        are studied by means of a covariant Faddeev equation.
        They are homogeneous integral equations for a hadron's amplitude 
        and depend on the dressed quark propagator as well as the quark-antiquark or three-quark kernel, respectively.

        The covariant Faddeev equation was recently solved for the nucleon mass by implementing a rainbow-ladder (RL) truncation,
        i.e. a dressed gluon-ladder exchange kernel between any two quarks~\cite{Eichmann:2009qa,Eichmann:2009en,Alkofer:2009jk}.
        While this puts the analysis of baryon properties on the same footing as sophisticated meson studies,
        the numerical efforts are involved. To simplify the problem, a quark-'diquark' bound-state BSE
        to study baryon properties has often been employed, see e.g.\ \cite{Oettel:1998bk}.
        It is based on the observation that the attractive nature of quark-antiquark
        correlations in a color-singlet meson is also attractive for $\bar{3}_C$ quark-quark correlations
        within a color-singlet baryon~\cite{Cahill:1987qr,Maris:2002yu}.

        In connection with hadronic bound-state equations,
        the RL truncation of DSEs has been widely employed for studying hadron observables.
        It provides a reasonable description of pseudoscalar-meson, vector-meson and
        nucleon ground-state masses and electromagnetic properties, see
        e.g.~\cite{Holl:2005vu,Maris:2005tt,Maris:2006ea,Bhagwat:2006pu,Eichmann:2007nn,Cloet:2008wg,Nicmorus:2008vb,Krassnigg:2009zh}.
        Other quantities, most notably the masses of axial-vector and pseudoscalar isosinglet mesons, are not reproduced so well. 
        Efforts to go beyond RL have been made (see e.g.~\cite{Alkofer:2008tt,Fischer:2009jm,Chang:2009zb}) but typically require a significant amplification of numerical effort.
        In recent studies
        a consistent implementation of additional structures in the quark-gluon vertex and quark-antiquark kernel has proven capable
        to provide a better description of such observables as well~\cite{Alkofer:2008et,Fischer:2008wy,Fischer:2009jm,Chang:2009zb,Chang:2010jq}.

        On the other hand, substantial attractive contributions beyond RL come from a pseudoscalar meson cloud which augments the 'quark core' of dynamically
	    generated hadron observables in the chiral regime, whereas it vanishes with increasing current-quark mass.
        A viewpoint explored in Ref.~\cite{Eichmann:2008ae} was to identify RL with the quark core of chiral effective field theory
        which, among other corrections, must be subsequently dressed by pion-cloud effects.
        Such a quark core can be modeled by a current-quark-mass dependent input scale which is deliberately inflated close to the chiral limit.
         Resulting mass--dimen\-sionful $\pi$, $\rho$, $N$ and $\Delta$ observables were shown to be consistently overestimated and mostly
         compatible with quark-core estimates from quark models and chiral perturbation 
	 theory~\cite{Eichmann:2008ae,Eichmann:2008ef,Nicmorus:2008vb},
	 a pattern also present in a recent exploratory study of the QCD chiral transition temperature in this
	 approach \cite{Blank:2010bz}.

        In the present work we follow this point of view 
        and report on the latest calculations
        of the quark-core contributions to the $\Delta$ mass and electromagnetic form factors.

\begin{figure}
                    \begin{center}
                    \includegraphics[scale=1.6]{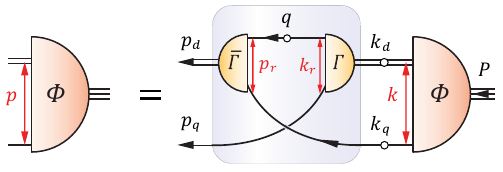}
                    \caption{(Color online) The quark-diquark BSE.\label{fig:1}}
                    \end{center}
                    
\end{figure}


\section{Quark-diquark Faddeev-equation framework} \label{sec:qdq}

       In a quark-'diquark' scenario, a color-singlet baryon emerges as a bound state of a color-triplet quark and  
       color-antitriplet diquarks. Diquark correlations are implemented as a separable sum
       of pseudopar\-ticle pole contributions in the quark-quark scattering matrix. 
       This procedure simplifies the covariant Faddeev equation to a quark-diquark BSE on the baryon's mass shell 
       which is diagrammatically represented in Fig.~\ref{fig:1}.

       The baryon mass and amplitude are obtained as numerical solutions of the quark-diquark BSE
       once all its ingredients are specified: the dressed-quark propagator (single line),
       the diquark propagator (double line),
       and the diquark amplitude $\Gamma^\nu$ and its charge-conjugate $\bar{\Gamma}^\nu$ which appear
       in the quark-diquark kernel. The binding mechanism in the baryon is realized via an
       iterated exchange of roles between the single quark and any of the quarks contained in the diquark.

       The fundamental building block which connects the properties of baryons
       with the underlying structure of QCD is the renormalized dressed quark propagator.
       It involves the quark mass function $M(p^2)$ which is non-perturbatively enhanced
       at small momenta and thereby indicates the dynamical generation of a large constituent-quark mass.
       This manifestation of dynamical chiral symmetry breaking emerges in the solution of the quark DSE.
       The latter involves the gluon propagator and the quark-gluon vertex that both satisfy their 
       own DSEs which depend on higher-order Green functions.
       In practical calculations, the resulting infinite set of coupled DSEs is circumvented
       by employing truncations that preserve the underlying symmetries of QCD.

       In connection with meson properties, e.g. to establish the pion as the Goldstone boson of spontaneous chiral symmetry breaking,
       it is imperative to preserve the axial-vector Ward-Takahashi identity.
       It connects the kernel of the quark DSE with that of the pseudoscalar meson BSE,
       ensures a massless pion in the chiral limit and leads to a generalized Gell-Mann--Oakes--Renner relation
       \cite{Maris:1997hd,Holl:2004fr}.   
       The simplest truncation that satisfies this constraint is the rainbow-ladder (RL) truncation
       which amounts to an iterated dressed-gluon exchange between quark and antiquark
       and has been extensively used in Dyson-Schwinger studies of hadrons, see e.g.~\cite{Krassnigg:2009zh,Eichmann:2007nn} and references therein.

       The RL truncation retains only the vector part of the dressed quark-gluon vertex.
       Its non-perturbative dressing, together with that of the gluon propagator, is absorbed into an effective coupling
       $\alpha(k^2)$ which represents the only unknown function of the model.
       Herein we employ for $\alpha(k^2)$ the frequently used ansatz~\cite{Maris:1999nt} 
                    \begin{equation}\label{couplingMT}
                        \alpha(k^2) = \pi \eta^7  \left(\frac{k^2}{\Lambda^2}\right)^2 \!\! e^{-\eta^2 \left(\frac{k^2}{\Lambda^2}\right)} + \alpha_\text{UV}(k^2) \,,
                    \end{equation}
                    where $k$ is the gluon momentum.
        At large gluon momenta, the second term in the effective coupling $\alpha(k^2)$ decreases logarithmically and reproduces
        QCD's perturbative running coupling. At small and intermediate gluon momenta, the first term 
        must exhibit sufficient strength to allow for dynamical chiral symmetry breaking and the dynamical generation of a
        constituent-quark mass scale. The infrared behavior of the effective coupling is controlled by two parameters:
        an infrared scale $\Lambda$ and a dimensionless width parameter $\eta$.

      The features of the effective coupling directly translate from the case of mesons 
      to that of diquarks which enter the quark-di\-quark bound-state equation in Fig.\,\ref{fig:1}.
      By virtue of the RL truncation, both mesons and diquarks are bound by the same gluon-exchange mechanism. 
      The corresponding diquark BSEs are obtained by assuming 
      timelike diquark poles at certain values of the total diquark momentum $P^2$ in the quark-quark scattering matrix, i.e.
      $P^2=-m_\text{sc}^2$, $P^2=-m_\text{av}^2$, which characterize the lightest diquarks, namely the scalar and axial-vector ones.
        Diquarks carry color and are hence not observable; yet such a pole structure does not contradict diquark confinement, see e.g.~\cite{Alkofer:2000wg}.
        In the present context, timelike diquark poles emerge as an artifact of the RL truncation which does not persist beyond RL~\cite{Bender:1996bb};
        nevertheless they indicate the presence of diquark mass scales within a baryon.
	Studies in a similar setup provide further support for diquark correlations as a reasonable concept for 
	the description of baryons \cite{Alkofer:2005ug}.
        While both scalar and axial-vector diquark correlations are important for the description of the nucleon, 
        the spin$-3/2$ and isospin$-3/2$ flavor symmetric $\Delta$ necessitates only axial-vector diquark correlations.

        The scalar and axial-vector diquark BSEs only specify the on-shell diquark amplitudes whereas 
        diquarks in a baryon are off-shell. 
        Within the separable ansatz for the scattering matrix, 
        information on its off-shell behavior is encoded in
        the scalar and axial-vector diquark propagators.
        By reinserting the separable pole ansatz into the Dyson series for the scattering matrix,
        the resulting diquark propagators are completely specified from their substructure; see \cite{Nicmorus:2008vb,Eichmann:2009zx} for details.


\section{Nucleon and $\Delta$ masses}
\label{sec:5}

        All the ingredients of the quark-diquark BSE are now determined: the quark propagator is obtained
        as a solution of the quark DSE, the scalar and axial-vector diquark amplitudes as solutions of the diquark BSEs,
        and the respective diquark propagators follow from the separable diquark-pole ansatz.
        All these elements enter the quark-diquark kernel in Fig.~\ref{fig:1} and are numerically calculated
        within RL truncation which involves only one parametrization as its input, namely the effective coupling $\alpha(k^2)$ of Eq.\,\eqref{couplingMT}.

\begin{figure}
                    \begin{center}
                    \includegraphics[scale=0.95]{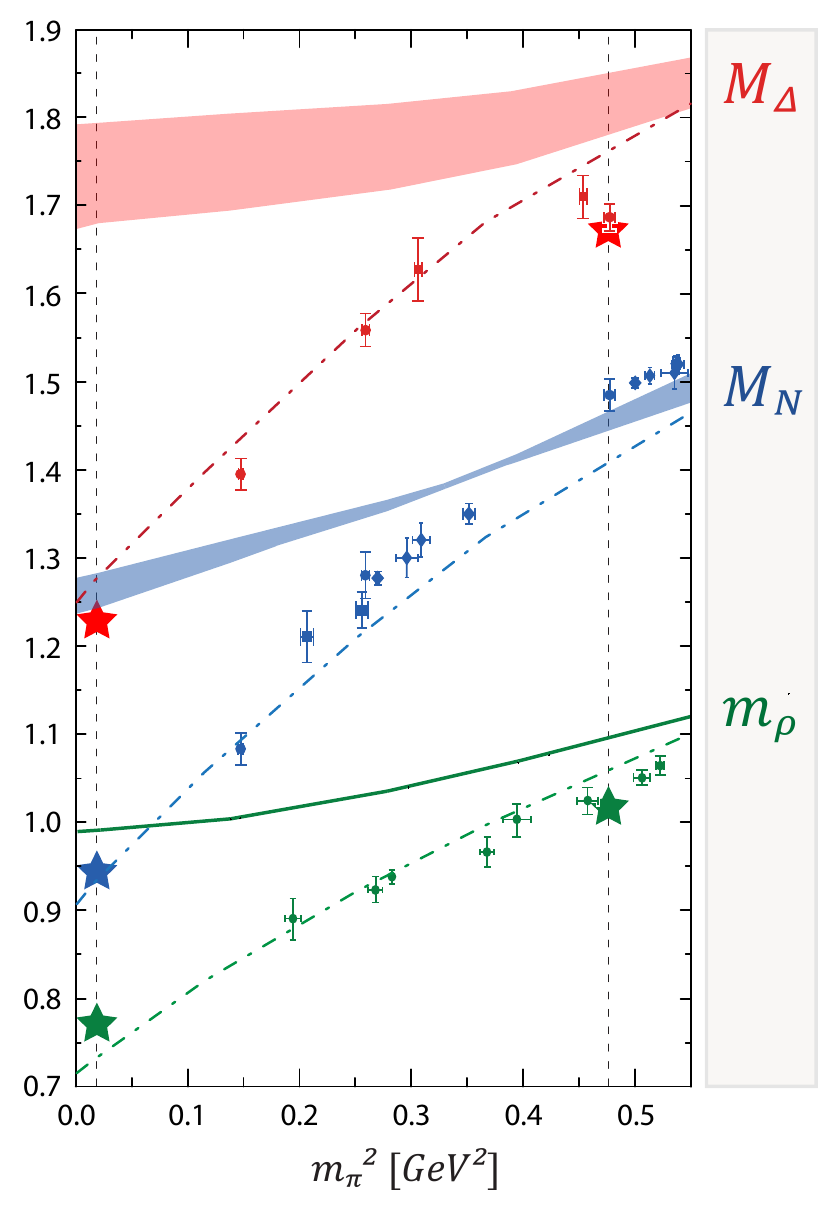}
                    \caption{\label{fig:2}Evolution of $\rho$-meson, nucleon and $\Delta$ masses with $m_\pi^2$ as obtained from their $q\bar{q}$ and quark-diquark BSEs.
			                 The bands correspond to the core model and represent the sensitivity of the masses to the width parameter $\eta = 1.8\pm 0.2$.
                             The dash-dotted lines depict the results in the fixed-scale model for the central value $\eta = 1.8$.
                                            Stars denote the experimental values~\cite{Amsler:2008zzb} of $\rho$, $N$ and $\Delta$ at the $u/d-$quark mass and $\phi$, $\Omega$ at the strange-quark mass whose positions are indicated by vertical lines.
                                            Note that there is no $s\bar{s}$ pseudoscalar meson in nature; the value $m_{s\conjg{s}}=0.69$ GeV corresponds to a meson-BSE solution
                                            at a strange-quark mass $m_s = 150$ MeV~\cite{Holl:2004fr}.
                             We compare with a selection of lattice data for
                             $m_\rho$~\cite{Ali_Khan:2001tx,Allton:2005fb}, $M_N$~\cite{Ali_Khan:2003cu,Leinweber:2003dg,Frigori:2007wa,Alexandrou:2009hs} and $M_\Delta$~\cite{Zanotti:2003fx,Alexandrou:2009hs}.}
 \end{center}
        \end{figure}

        Upon a decomposition of the quark-diquark amplitudes of nucleon and $\Delta$ into orthogonal sets of Dirac covariants,
        their amplitudes and masses emerge as numerical solutions of the respective quark-diquark BSEs
        (details on the calculation were reported in Ref.~\cite{Nicmorus:2008vb}).
        We depict the results for nucleon and $\Delta$ masses, together with that of the $\rho$-meson, in Fig.~\ref{fig:2} and compare
        their evolution with $m_{\pi}^2$ to lattice calculations.
        The masses are calculated within two versions of the model which are characterized by the infrared scale $\Lambda$ in the effective coupling of Eq.\,\eqref{couplingMT}.
        The dashed lines in Fig.\,\ref{fig:2} are obtained by
        using a fixed scale $\Lambda=0.72$ GeV which is
        adjusted to reproduce the experimental pion decay constant and kept fixed for all values of the quark mass.
        It yields the result $M_N=0.94$ GeV and $M_\Delta=1.28$ GeV at the physical $u/d$-quark mass 
        corresponding to a pion mass $m_\pi=140$ MeV. 
        These results are reasonably close to the experimental values and consistent with
        pseudoscalar-meson and vector-meson ground-state properties which are satisfactorily reproduced in this setup,
        e.g.\ \cite{Maris:2006ea,Krassnigg:2009zh,Nicmorus:2008vb},
        and moreover insensitive to the shape of the coupling in the infrared,
        i.e. to a variation of the width parameter $\eta$~\cite{Maris:1999nt}.

        In the second version of the model, the hadronic quark-core properties are implemented  
        through a current-quark mass dependent scale $\Lambda$ in Eq.\,\eqref{couplingMT}.
        It is deliberately inflated close to the chiral limit
        and fixed to reproduce the core properties of the $\rho$ meson~\cite{Eichmann:2008ae}.
        The corresponding results in Fig.\,\ref{fig:2} are depicted by bands which indicate
        the variation with the width parameter $\eta$.
        For a value of $\Lambda=0.98$ GeV at the $u/d$ mass, the resulting values are $M^{core}_N=1.26(2)$ GeV and $M^{core}_\Delta=1.73(5)$ GeV.
        In the chiral region, the core version of the model uniformly overestimates the experimental and lattice data as well as the results
        obtained using the fixed-scale model,
        whereas this deviation decreases with increasing current-quark mass. 
        While pionic corrections to hadronic observables vanish at large quark masses,
        they should reduce the core masses of nucleon and $\Delta$ by $\sim 300$ MeV in the chiral region.
        In this respect, the core value for $M_N$ is roughly consistent with a
	    pseudoscalar-meson dressing providing the dominant correction to the quark-diquark
	    core, whereas the somewhat large deviation between the experimental and 'core' mass
        of the $\Delta$ may indicate the relevance of further diquark channels in describing $\Delta$ properties.

        We note that simple relations between the two setups hold in the chiral limit~\cite{Eichmann:2009zx} 
        where the infrared parameter $\Lambda$ in Eq.\,\eqref{couplingMT} represents the only relevant scale in the system, i.e. the scale of dynamical symmetry breaking.
        This implies that mass-dimensionful quantities which are sensitive to the infrared properties scale with $\Lambda$.
        For a given set of dimensionful observables, the fixed-scale model produces results 
        that are overestimated by the same percentage upon entering its core version.
        On the other hand, the distinction between dimensionless quantities calculated
        within either versions of the model becomes irrelevant in the chiral region.
        As we will point out below, this is the case for electromagnetic form factors.

\section{Delta electromagnetic form factors}
\label{sec:6}

       To compute the electromagnetic form factors of the $\Delta$-baryon, one must specify how the photon
       couples to its constituents. In the quark-diquark context this amounts to resolving the coupling of the photon
       to the dressed quark and the axial-vector diquark (impulse approximation), to the axial-vector diquark amplitude (seagulls) 
       and to the exchanged quark in the quark-diquark kernel. With this decomposition the $\Delta$ electromagnetic current is automatically conserved~\cite{Oettel:1999gc}.

       At the level of the constituents, current conservation 
       translates to electromagnetic Ward-Takahashi identities (WTIs) which 
       constrain the longitudinal parts of the above vertices and unambiguously relate them to the previously determined
       quark and diquark propagators and diquark amplitudes.
       On the other hand, current conservation only partly constrains pieces which are transverse to the photon momentum.
       Their important role in physical observables is accounted for by
       augmenting the vertices as determined from their WTIs by appropriate transverse $\rho$-meson pole contributions. 
       The details of the construction of the electromagnetic current are presented in Appendix~C of~\cite{Nicmorus:2010sd}.

\begin{figure*}
                   \begin{center}
                    \includegraphics[scale=1.5]{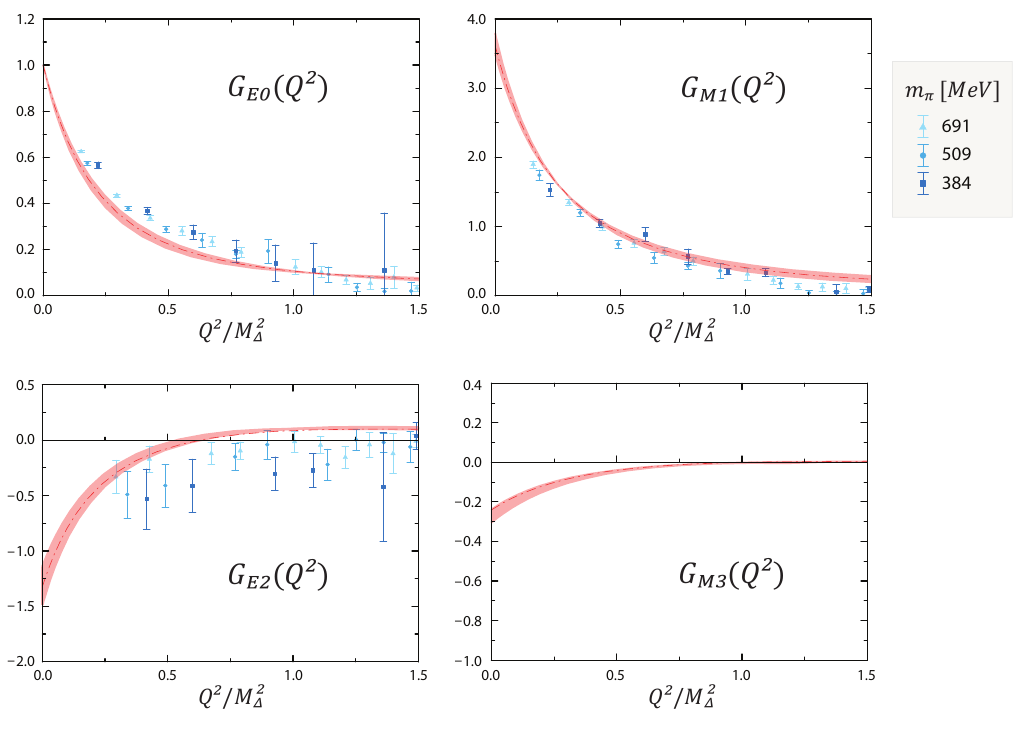}
                   \caption{\label{fig:3}(Color online) Electromagnetic form factors of the $\Delta$ calculated at the physical point $m_\pi=140$ MeV, and
                                           compared to unquenched lattice data of Ref.~\cite{Alexandrou:2009hs} at three different pion masses.
                                           The bands represent the sensitivity to a variation of the infrared width parameter $\eta=1.8\pm 0.2$.
					   Adapted from Ref.~\cite{Nicmorus:2010sd}}
				\end{center}
        \end{figure*}

       Following these prescriptions for the electromagnetic current, the $\Delta$ electromagnetic form factors ---
       the electric monopole $G_{E0}(Q^2)$, electric quadrupole $G_{E2}(Q^2)$,
       magnetic dipole $G_{M1}(Q^2)$ and magnetic octupole form factor $G_{M3}(Q^2)$ ---
       are directly related to the effective quark-gluon coupling in Eq.\,\eqref{couplingMT}.
       In Fig.~\ref{fig:3} we depict the 'core' contributions to the $\Delta^+$ electromagnetic form factors and compare to recent lattice data~\cite{Alexandrou:2009hs}.
       Due to isospin symmetry the $\Delta^{++}$, $\Delta^0$ and $\Delta^-$ form factors
       are simply obtained by multiplying those of the $\Delta^{+}$ with the appropriate charges.
       The electromagnetic form factors (as dimensionless quan\-tities) are plotted
       as a function of the dimensionless variable $Q^2/M_\Delta^2$ for our data, and $Q^2/(M_\Delta^\text{lat})^2$
       for the lattice data. This enables an unambiguous comparison between our form factor results,
       calculated with an implicit 'core' mass $M_\Delta>M_\Delta^\text{exp}$ and the lattice results.
       From another perspective, the calculated $M_\Delta$ sets the scale of dynamical chiral symmetry breaking
       in either version of the model, and hence the dimensionless form factors in Fig.~\ref{fig:3}
       would approximately match the corresponding dimensionless quantities calculated within the fixed-scale model.

       Experimentally only the magnetic moments of the $\Delta^{+}$ and $\Delta^{++}$ are known,
       albeit with large errors. For instance, for the $\Delta^{+}$ the  Particle Data Group quotes the value
       $3.5^{+7.2}_{-7.6}$~\cite{Amsler:2008zzb}. Our result for the magnetic moment,
       $G_{M1}(0) = 3.64(16)$, compares well with quark-model predictions 
       and chirally extrapolated lattice results; see~\cite{Ramalho:2010xj} and references therein.
       The deformation of the $\Delta$-baryon is encoded in its electric quadruple and magnetic octupole moments.
       Currently there are no experimental measurements for these observables.
       From the measurement of the $N\gamma\Delta$ transition, one can extract
       the value $G_{E2}(0)=-1.87(8)$ in the large-$N_C$ limit~\cite{Buchmann:2002mm,Alexandrou:2009hs};
       similar values are predicted by constituent-quark models~\cite{Ramalho:2009vc}.
       Lattice calculations indicate a negative value for $G_{E2}(0)$ as well~\cite{Alexandrou:2009hs} but are limited by large statistical errors.
       Our result for the electric quadrupole moment, $G_{E2}(0)=-1.32(16)$,
       is negative and compatible with these observations. We note that $G_{E2}(Q^2)$ develops a zero-crossing at $Q^2/M_\Delta^2\sim 0.6$,
       a feature which is unexpected but not clearly excluded from the available lattice results.
       Our calculation for the magnetic octupole moment yields a small and negative value $G_{M3}(0)=-0.26(4)$.
       We note that the electric quadrupole and magnetic octupole form factors are negative throughout the
       current-quark mass range
       which indicates an oblate deformation of the $\Delta$'s charge and magnetization distributions.

       We conclude that the rainbow-ladder truncated Poin\-car\'{e}-covariant Dyson-Schwinger/Bethe-Salpeter setup,
       upon implementing an appropriate input scale, produces consistently overestimated core contributions to nucleon and $\Delta$ masses. The overall attractive effect of chiral corrections is expected to shift the core masses to the experimental values. A $20-30$\% reduction for dynamically generated hadron masses in the chiral region is anticipated, while pionic effects decrease at larger quark masses.

       The impact of chiral corrections upon the low-$Q^2$ behavior of the $\Delta$ electromagnetic multipole form factors remains at present unclear.
       While near-future measurements at MAMI and JLab facilities remain to validate our predictions for the $\Delta$'s electromagnetic properties,
       the results collected herein show good agreement with lattice results and quark model analyses.
       Towards a complete understanding of the structure of baryons, our approach can be improved by
       implementing chiral corrections to the rainbow-ladder truncation, and augmented by studies which eliminate the diquark ansatz
       in support of a fully Poincar\'{e}-covariant solution of the three-quark Faddeev equation. 
       Fruitful insight can be gained by a forthcoming extension of our approach
       to the investigation of the $N\Delta\gamma$ transition.


\begin{acknowledgements}

     We thank
     M.~Blank, I.\,C.~Clo\"et, C.\,S.~Fischer, G.~Ramalho, M.~Schwinzerl, and R.~Williams
     for fruitful discussions. This work was supported by the Austrian Science Fund FWF under
     Projects No.~P20592-N16, No.~P20496-N16, and Erwin-Schr\"odinger-Stipendium No.~J3039,
     by the Helmholtz Young Investigator Grant VH-NG-332, and
     by the Helmholtz  International Center for FAIR within the framework of the LOEWE program launched by the State of Hesse,
     GSI, BMBF and DESY.
\end{acknowledgements}


\bigskip


\bibliographystyle{spbasic-mod}      



\end{document}